# *Electronic interaction of slow hydrogen and helium ions in the nickel - silicon system*

Tuan Thien Tran[1], Lukas Jablonka[2], Barbara Bruckner[1,3], Stefanie Rund[3], Dietmar Roth[3], Mauricio A. Sortica[1], Peter Bauer[1,3], Zhen Zhang[2], Daniel Primetzhofer[1]

[1] *Department of Physics and Astronomy, Ångström Laboratory, Uppsala University, Box 516, SE-751 20 Uppsala, Sweden*

[2] *Solid State Electronics, The Ångström Laboratory, Uppsala University, SE-75121 Uppsala, Sweden*

[3] *Johannes-Kepler University Linz, IEP-AOP, Altenbergerstraße 69, A-4040 Linz, Austria*

Corresponding author: Tuan Tran (tuan.tran@physics.uu.se)

Electronic stopping cross sections (SCS) of nickel, silicon and nickel-silicon alloys for protons and helium (He) ions are studied in the regime of medium and low energy ion scattering, i.e., for ion energies in the range from 500 eV to 200 keV. For protons, at velocities below the Bohr velocity the deduced SCS is proportional to the ion velocity for all investigated materials. In contrast, for He ions non-linear velocity scaling is observed in all investigated materials. Static calculations using density functional theory (DFT) available from literature accurately predict the SCS of Ni and Ni-Si alloy in the regime with observed velocity proportionality. At higher energies, the energy dependence of the deduced SCS of Ni for protons and He ions agrees with the prediction by recent time-dependent DFT calculations. The measured SCS of the Ni-Si alloy was compared to the SCS obtained from Bragg's rule based on SCS for Ni and Si deduced in this study, yielding good agreement for protons, but systematic deviations for He projectiles, by almost 20%. Overall, the obtained data indicate the importance of non-adiabatic processes such as charge-exchange for proper modelling of electronic stopping of in particular medium-energy ions heavier than protons in solids.

# Introduction:

An ion moving within a medium interacts with both the electrons and the nuclei of the material and thus gradually dissipates its kinetic energy. Studying this deceleration of ions in matter, commonly denoted the stopping process, is highly relevant for the fundamental understanding of the ion-matter interaction as well as for a series of technical applications such as space weathering, development of wall materials for plasma systems, cancer therapy and particularly for establishing accurate composition depth profiles in materials research [1,2]. An experimentally accessible key quantity in studies of ion-solid interaction is the stopping power $S$, which is the energy loss of the ion per unit path length: $S = \mathrm{d}E/\mathrm{d}x$. As the specific material density, in particular at the nanoscale might deviate from bulk atomic densities $n$, the stopping cross section (SCS), $\varepsilon = (1/n) \cdot (\mathrm{d}E/\mathrm{d}x)$, is commonly employed.

For H and He ions at kinetic energies of at most several keV, nuclear interactions contribute significantly to the energy loss experienced along the average ion trajectory, due to the large scattering cross sections [3]. At kinetic energies > 10 keV electronic interactions typically dominate the observed energy loss. For fast ions in solids (several MeV), electronic stopping is mainly due to excitation and ionization of core shell electrons. In this regime, the stopping power can be well described by the first principle theories [4]. For ion energies around 100 keV/u and below, the valence electrons and details in the density of states become more important. With velocities less than the Bohr velocity, $v < v_0$, the valence electrons have often been described as a free electron gas (FEG). Within this model, Fermi and Teller [5] and later refined studies [6-8] have predicted velocity proportionality of the stopping power, $S = Q(Z_1, r_S) \cdot v$, where $Q$ can be referred to as the friction coefficient of the FEG. The $Q$ values depend on the atomic number of the projectiles ($Z_1$) and the Wigner-Seitz radius of the FEG, $r_S$ [5]. Good agreement with experimental results has been obtained when $Q$ was calculated by static DFT for a FEG with an effective density, $r_S$, deduced from experimental plasmon energies [9].

Proportionality between the stopping power and the ion velocity has been experimentally observed in many materials for protons at $v < v_0$ [10]. However, deviation from $S \propto v$ has also been observed and traced back to band structure effects and to charge exchange processes [11-14]. For example, the friction coefficient $Q$ for proton in gold (Au) was found to decrease significantly for ion velocities < 0.6 a.u. [11], due to the excitation threshold of the d-band. Consequently, ions are expected to only excite electrons in the s-band at sufficiently low velocity. Similar behavior was later observed for copper and silver [12]. Electronic energy loss was found to vanish in large band gap materials, such as lithium fluoride (LiF) [13]. At ion velocities $< 0.1$ a.u. (250 eV/u), ions were found to travel through the material without transferring energy to the electronic system. Note, that the observed threshold velocities are much lower than expected from the binary collision model.

Additional complexity is observed for ions with higher atomic numbers, such as helium, in simple metals like Al: while at $v < v_F$ proton stopping is perfectly velocity-proportional [14] and in perfect agreement with DFT predictions, for He a non-linear behavior is observed, which cannot be explained in terms of the band structure of Al. A similar behavior has also been observed for other target materials. Complementary studies using Auger electron spectroscopy also indicate that for He the formation of molecular orbitals leads to excitation of inner shells, thereby opening additional energy dissipation channels [15]. Also recent calculations by time-dependent DFT (TD-DFT) indicate the importance of dynamic phenomena for the description of the energy loss at $v < v_F$ [16,17].

Commonly, in applications, the interest is focused on compound targets. The SCS of a compound $A_xB_{1-x}$ can be estimated according to Bragg's rule (stopping power additivity), $\varepsilon_{A_xB_{1-x}} = x \cdot \varepsilon_A + (1-x) \cdot \varepsilon_B$, where $x$ is the atomic fraction of element A [18]. This prediction is commonly employed and proves highly accurate at energies of several MeV/u [19]. However, due to the above mentioned increasing relative weight of the valence electrons at low-to-medium energies, chemical binding and the physical phases of the materials have been found to affect the stopping power considerably [20,21]. This fact obviously challenges the reliability of predictions of stopping powers by software packages such as SRIM [22], which rely on extrapolation from often limited experimental datasets [23,24].

In this contribution, we will present measurements of electronic excitations in nickel, silicon and nickel silicide, Ni-Si, by H and He ions in the energy range from $\sim 0.5$ keV to $\sim 200$ keV. This choice of target materials permits to study the validity of the FEG model and to look for chemical effects in the compound. Via providing benchmark data in a regime in which velocities of valence electrons and the intruding particle are comparable this study is also relevant for theoretical models for electron dynamics in solids. Furthermore, these data are highly relevant for technological applications as the materials are key components in present electronic devices.

**Experimental details:**

Two distinct, non-overlapping energy regimes were explored using two different set-ups and two sets of samples. For ion energies above several tens of keV, the following targets were used: (1) ~20 nm polycrystalline Ni on ~95 nm $SiO_2$/(100) Si as substrate, (2) ~20 nm Si on glassy carbon and (3) ~20 nm Ni silicide film on ~95 nm $SiO_2$/(100) Si. The specific choice for the substrates was made to maximize energy separation between backscattering from the films and the substrates in the spectra. The $SiO_2$ layers were prepared by plasma-enhanced chemical vapor deposition. The Ni films in sample (1) and (3) was prepared by magnetron sputtering (von Ardenne CS730S, 150 W, $6 \times 10^{-3}$ mTorr Ar atmosphere, deposition rate 30 nm/min). The Si films in samples (2) and (3) were prepared by the same tool with different deposition parameters (250 W, $6 \times 10^{-3}$ mTorr Ar atmosphere, deposition rate 17 nm/min). For sample (3), after the subsequent depositions of the Si and the Ni films the nickel silicide layer was formed by annealing the sample with rapid thermal processing at 450 ℃ for 30 s in $N_2$ atmosphere (ramp rate 10 ℃/s). The unreacted Ni film was subsequently removed by a mixture of $H_2SO_4$ and $H_2O_2$.

Accurate measurements of the areal thicknesses of the films were performed by Rutherford backscattering spectrometry (RBS), using 2 MeV $He^+$ provided by the 5 MV 15SDH-2 tandem accelerator at Uppsala University. Two measurements at different geometries were performed on each sample: azimuth angle 5°/tilt angle 2° and azimuth angle 40°/tilt angle 2°. The specific angles were chosen to minimize channeling effects that could introduce errors in the thickness measurement, which would directly be propagated in the deduced SCS. Additional suppression of possible channeling effects was done with multiple small random angular movements around the equilibrium angles within a range of 2°. The obtained areal thickness values for the two different geometries are found to differ by less than < 3%. In order to reduce possible systematic errors in the $charge \times solid\ angle$ product from insufficiently well-known stopping powers in the glassy carbon substrate, additional relative stopping power measurements [25,26] have been performed to confirm the thickness of the Si film in sample (2). Assuming bulk densities of the films results in average thicknesses of the Ni, Si and Ni silicide films of 21.9 nm, 22.4 nm and 22.4 nm, respectively. RBS characterization also indicated a certain degree of oxygen and argon contaminants in the films. Quantification of the oxygen contamination in the films was done with elastic scattering using the nuclear resonance between O and 3.04 MeV He ions with a drastically enhanced the scattering cross section of O [27]. In summary, the composition and the amount of contamination in the films is determined as follows: sample (1): 95.3% Ni/4.7% O, sample (2): 89% Si/5% O/6% Ar and sample (3): 62% Ni/35% Si/1% O/2% Ar.

For determining electronic SCSs, time-of-flight spectra of ions with primary energies in the range of 25 keV − 200 keV backscattered from the films of interest were measured by the time-of-flight medium-energy-ion-scattering system (MEIS) at Uppsala University. Details of the set-up are described elsewhere [28]; here we summarize only the most relevant features. The stop detector of the system is a large-area delay line detector (DLD120) from Roentdek. It is positioned at a distance of ∼290 mm from the samples, with a solid angle of 0.13 sr. This large area detector is able to acquire backscattering spectra with sufficient statistics for comparably low primary beam doses (at most several tens of nC). Therefore, the measurement is unlikely to induce any noticeable damage or sputtering to the films. Note, that for acquisition of those spectra to be finally evaluated only a limited section of the detector corresponding to a scattering angle ±2° degrees is employed, with the above mentioned advantages being unaffected.

In the LEIS regime, energy loss measurements were also performed at the time-of-flight low-energy-ion-scattering (ToF-LEIS) setup ACOLISSA, which was located at the University of Linz [29]. Atomic as well as molecular ion beams with a primary energy between 0.5 keV and 10 keV are used and projectiles backscattered into the stop detector located at an angle of $\theta$ =129° are detected. Three different sets of samples have been investigated: (i) in-situ grown Ni films with 9.1 nm, 8.3 nm and 6.5 nm thickness on a thick B film on Si [30], (ii) a high purity polycrystalline Ni sheet and (iii) thick amorphous Si. The thicknesses of the Ni films were determined via RBS by the use of an AN700 van de Graaf accelerator at the University of Linz.

For evaluation of electronic SCSs from the experimental spectra Monte-Carlo (MC) simulations using the TRBS (TRIM for Backscattering) code were employed [31]. This code permits to choose between different tunable screened scattering potentials and calculates the energy spectra of backscattered ions considering electronic stopping and multiple scattering [32], which is important at low ion energies. In our evaluation, the Ziegler-Biersack-Littmark (ZBL) potential was used as it shows to properly produce the scattering background caused by multiple scattering. In the MC the electronic SCS can be varied via a multiplicative factor to the predictions by SRIM. Alternatively, a table with reference values can be used. Thus, our SCS values are determined by fitting the simulated spectra to the experimental ones, by optimizing the SCS in the simulation. Fig. 1 shows two different backscattering spectra. The data shown in Fig. 1 (a) is obtained for 100 keV He$^+$ ions scattered from the 21.9 nm Ni thin film sample and presented together with three calculated spectra obtained by TRBS. The simulation with optimized SCS (red curve) reproduces the experimental spectrum very well, including the background yield due to plural and multiple scattering located between the signals associated with single scattering from Ni and Si (48 keV − 60 keV). The dashed lines in Fig. 1 represent simulation results obtained for slightly different SCS values (optimized SCS ±4%) to demonstrate the sensitivity of our evaluation procedure. To show the validity of our approach for the whole range of investigated energies, Fig. 1(b) shows a typical spectrum obtained at low energies, i.e. for 5 keV He$^+$. Again, simulated data is shown, similar as for (a) [33]. Experimental statistics have only minor impact on the accuracy of the obtained results, a systematic uncertainty of about ±3% can be expected in the resulting stopping power, with the major source of uncertainty constituting of the thickness calibration of the RBS [34]. Note that an error in the thickness would affect only the absolute value of the SCS, but not its velocity dependence. In particular, the ratio of the SCS for He and H would remain unaffected.

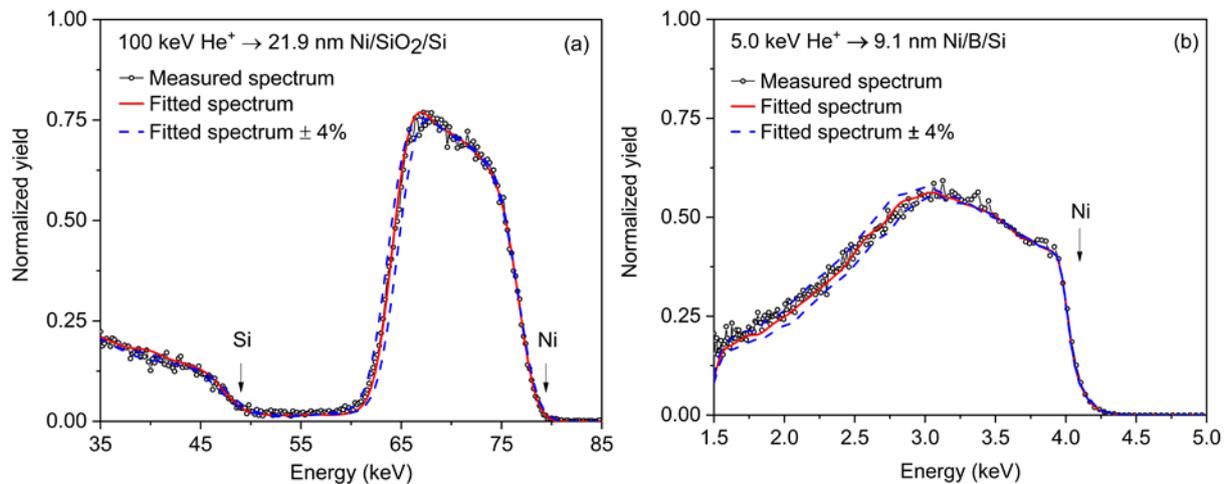

Fig. 1: Experimental spectrum for 100 keV (a) and 5 keV He$^+$ (b) primary ions scattered from the Ni samples. Beside the best fit (red curve), two other simulated spectra with the electronic SCS modified by ±4% are added to provide an assessment of achievable precision in the stopping power measurement.

In the LEIS regime the thin Ni films are analyzed as described in the paragraph above, where the energy loss is obtained from fitting the width of the Ni peak by tuning the electronic energy loss in the simulations. For the low-energy data for bulk Si, and several data points obtained from the high-purity Ni sheet, electronic stopping is evaluated from the ratio of fitted spectra reproducing the height ratio of the energy spectra close to the leading edge for the sample of interest and a reference sample [25]. Again, MC simulations were employed to consider multiple scattering contributions. In order to have similar scattering cross sections and therefore multiple scattering contributions in both reference and material of interest, we used Cu [35] and Al [14] as references for Ni and Si, respectively. Excellent agreement of data obtained independently by both methods is observed, as earlier reported for both protons and He ions for a range of target materials [30,42].

**Experimental results**

In Fig. 2, the experimentally deduced SCS values of Ni for H and He ions are presented. The figure holds also a selection of previously published experimental data [30,36-40], predictions using the stopping and range of ion in matter (SRIM) [22] as well as DFT results [41]. For H ions (Fig. 2a), our results (red squares) cover the range of SCS values slightly below the stopping maximum (Bragg's peak), with excellent agreement with SRIM, which in turn represents a good approximation to the majority of published data (see also [42]). At $v < 1$ a. u., our data is converging towards velocity proportionality, as one may expect for a metal with a d-band that extends up to the Fermi energy. For He ions, in Fig. 2b data measured by MEIS (red squares) and by LEIS (red circles) are presented. Unlike for H, in the SCS data for He ions a slight non-linearity appears, e.g., by the kink at the velocity of 0.2 a. u. and the offset of the MEIS data with respect to the LEIS data. Note that for platinum, a similar result has been reported [43], which has a density of states similar to Ni, both belonging to the same transition metal group in the periodic table. For H ions the SCS showed velocity proportionality as for a FEG material. For He ions, however, the SCS exhibits a kink at a velocity of 0.2 a. u.. Energy-dissipative processes different from direct electron-hole pair excitation in binary collisions such as charge-exchange can explain this non-trivial energy dependence of the He stopping power.

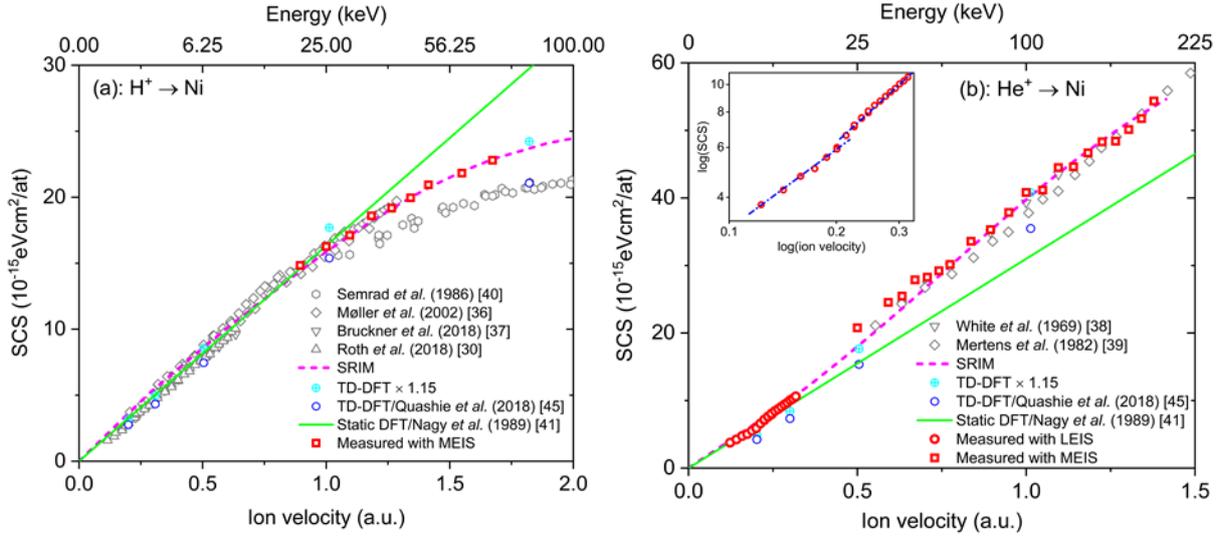

Fig. 2: Electronic stopping cross-section of Ni for the H ions (a) and He ions (b) as a function of the ion velocity. The figures also hold the predictions of the SCS using the DFT calculation of the FEG model.

We used the DFT calculation by Nagy *et al.* for a non-interacting electron gas [41], which are presented as the green lines in all figures. To find the predicted friction coefficient $Q$, we use the $r_S$ values deduced from experimental plasma frequencies, which are 1.8 a. u. and 1.97 a. u. for Ni and Si, respectively [44]. As one can see in Fig. 2, the DFT calculation agrees very well with our results for H stopping. Even if Ni cannot be regarded as a FEG material, ~ 3 electrons per atom may be taken as an effective number of free electrons for both, plasmon excitation by electrons and electronic stopping of ions. For He ions (see Fig. 2b) the DFT calculation appears to underestimate the SCS. This observation indicates that either electron-hole pair excitation in binary collisions by He has to be modelled differently, or, for He ions, an additional mechanism contributes to the energy loss beside the electron-hole pair excitation. In this context it is worth noting that a very recent calculation using time-dependent DFT [45] in Fig. 2 is able to reproduce the different velocity scaling of proton and helium stopping. For H ions, the TD-DFT values closely follow the measurement even beyond the proportional regime, and very well reproduce the velocity dependence of our data for He ions in the whole energy range investigated. Only the absolute values of TD-DFT and experiment differ by a factor of 1.15, both for H and He ions with calculations underestimating the experimental results, but well reproducing the observed change in the velocity dependence.

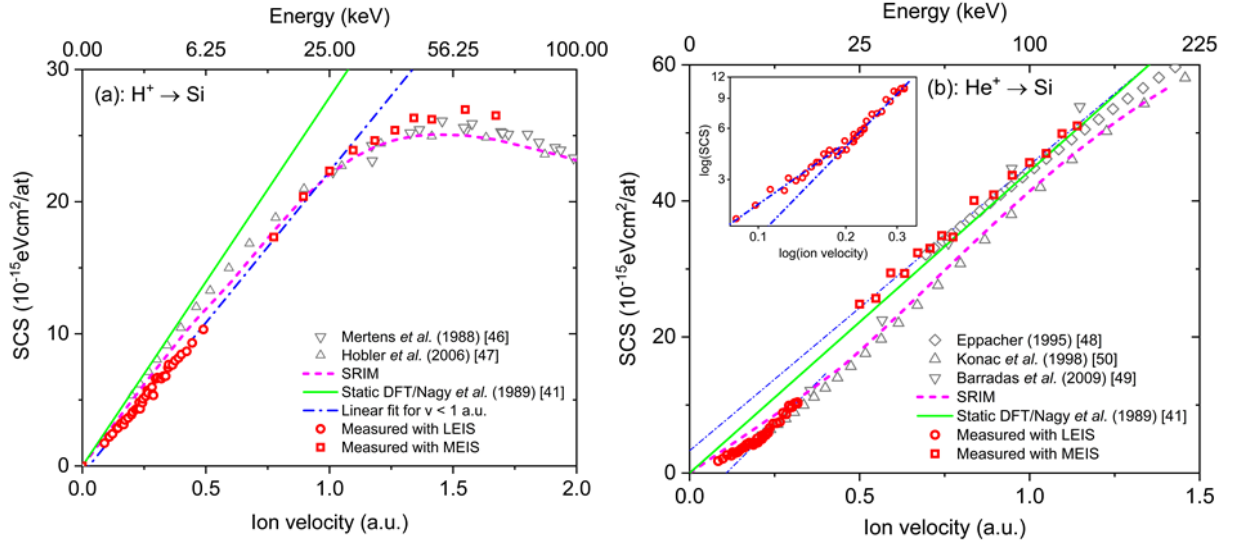

Fig. 3: Stopping cross-section of Si for the H ions (a) and He ions (b) as a function of the ion velocity. Our result includes the data from the measurement using MEIS (red squares) and independent study using LEIS (red circles). Some relevant references are also included [46-50].

The SCS results of Si for H and He ions are presented in Fig. 3, where data measured by MEIS (red squares) and by LEIS (red circles) are included. The data for H ions shows a linear velocity dependence. A fit to the low-energy data predicts a small but finite threshold for vanishing energy loss. This observation is well in accordance with the observation of larger thresholds for insulators [13] and comparably small values for semiconductors such as Ge [25]. In contrast to the observation for Ni, the SCS of Si is found lower than the DFT prediction. This observation can again be related to the fact that Si is a semiconductor with a bandgap of 1.1 eV. From the DFT model for non-interacting electrons [41] one would expect ~2 electrons per Si atom participating in the stopping process, as compared to about 4 electrons in the plasmon loss experiments. This disparity may be related to differences in the electronic excitation process by electrons and by H ions such as dynamic screening of the intruding ion.

The stopping data of the He ions are shown in Fig. 3(b). For this data set significant deviations from velocity-proportionality are observed. Extrapolation of the stopping data (see dash-dotted line in Fig. 3b) for $v > 0.5$ a. u. (25 keV) to lower velocities results in a positive intercept at $v = 0$, with a value of $3 \times 10^{-15}$ eV cm$^2$/at. Obviously, this extrapolation is unphysical, but rather suggests an additional energy loss channel for a He ion in Si, besides electron-hole pair excitation in a binary Coulomb collision. This additional process is expected to set in at a positive threshold velocity. This is why the SCS data in the interval 0.2 to 0.32 a. u. (4 − 10 keV) extrapolates toward a finite velocity of ~0.1 a. u. at $\varepsilon = 0$ (as indicated in the inset). Finally, at velocities < 0.2 a. u., the SCS is again found proportional to the ion velocity. A similarly complex velocity scaling of the electronic SCS has been observed previously in Al at a comparable range of ion velocities and was attributed to the charge exchange effect between Al and He along the trajectory of the He ions [14]. The equilibrium charge state of He traversing a solid at $v < 1$ a. u. can be expected to be close to 0 [51]. However, strong promotion of the He $1s$ level is known to occur at interaction distances He-Al below 0.75 a. u., leading to charge exchange processes [52]. Even at low velocities neutral He projectiles can be re-ionized in a sufficiently close collision, due to level promotion, on the expense of its kinetic energy. The same type of processes are also active in Si, due to the similarity between Si and Al in their electronic structures and their threshold energies for ionization of He$^0$ ($\sim 300$ eV) [53]. While for these processes to occur, at low energies, large scattering angles are required, with increasing energy these processes become more likely even for small angle scattering making them contribute to a larger fraction of possible trajectories [54]. The cross section for reaching a certain minimum interaction distance strongly increases with energy, in parallel with large angle scattering drastically decreasing in probability. In other words, these impact parameter dependent processes clearly introduce a trajectory dependence of the observed energy loss. However, at energies associated with the MEIS regime large angle scattering can be considered comparably rare along an average trajectory, and thus, as shown by Bauer and Mertens [55], only a small influence of the experimental approach on the deduced stopping cross section can be expected.. As discussed in Ref. [54] and [56] also at energies below 10 keV, scattering by small angle is dominating large angle scattering considerably, and the trajectory dependent processes are observable only in a very limited impact parameter range. These effects in combination are expected to render the stopping cross sections deduced from typical trajectories in the present approach valid also for trajectories typical for different experimental geometries, and thus results in the agreement observed in e.g. [30]. Similarly, at the present energies effective formation of molecular orbitals can be expected adding possible channels for energy dissipation as a function of energy [15,57]. A relative difference of these contributions, such as the different energy thresholds for charge exchange processes can explain the specific magnitude of the observed non-linearity for different systems. The DFT prediction (green line) appears to agree well with the measured SCS at $v > 0.5$ a. u. However, this observation may be purely coincidental as DFT calculations for He ions usually tend to under-estimate the SCS of FEG materials as shown in the case of Ni (Fig. 2(b)).

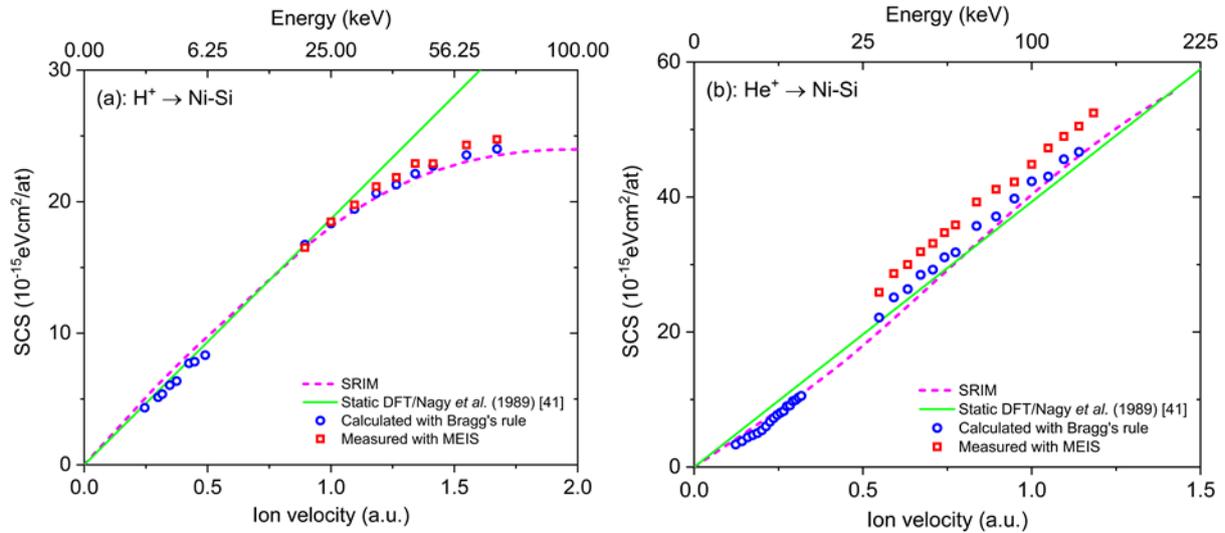

Fig. 4: Stopping cross section of the Ni-Si alloy as measured by MEIS (red squares) and as calculated from the SCS of the individual constituents using the Bragg's rule (dark-blue circles).

The final data set of our study comprises to the best of our knowledge the first experimental data set for the SCS of a nickel silicide for H ions and He ions. The RBS measurement showed the composition of this layer is 64%Ni and 36%Si, which is close to the common $Ni_2Si$ phase of the nickel silicide system. The red squares in Fig. 4 represent the data derived from MEIS experiments, whereas the dark-blue circles are calculated employing Bragg's rule, i.e., stopping power additivity [18], using the SCS of the individual constituents deduced in this study. For H ions, the SCS values of the measurements and the Ni – Si mixture are quite close to each other, with differences below 3%, the measured values being higher. At ion energies below 20 keV, measured SCS and Bragg's rule data agree within 1%, i.e., well within experimental uncertainties. In contrast, for He ions the measured SCS values are consistently higher than the SCS of the mixture (Bragg's rule, see Fig. 4b) by 8% at the highest energy up to 17% at the lowest energy. Deviations from the Bragg's rule are observed in many studies [20]. The effect is commonly found more pronounced at lower ion energies and in compounds of low atomic mass elements and strong polarity such as oxides, chlorides and fluorides. This behaviour can be understood on the basis of the modified electronic configuration of the compounds as compared to the pure elements. Silicides are not exhibiting a strong covalent character in their binding – indeed they are metallic in many aspects and exhibit very minor difference between proton stopping in the compound and in the mixture, while it shows a significant deviation for He.

Again, DFT predictions are done using the calculation by Nagy *et al.* and the $r_S$ value from the measured plasmon energy. For nickel silicide, studies have shown that the plasmon energies depend on the phases of the silicide, such as $Ni_2Si$, $NiSi$ and $NiSi_2$ [58-60]. Since the composition of our silicide film, after subtracting the percentage of the contaminants, is 64%Ni:36%Si, a plasmon energy of the $Ni_2Si$ phases is used, which is ~21.8 eV [58-60], corresponding to $r_S \approx 1.68$ a. u. As shown in Fig. 4a, the DFT prediction is in excellent agreement with our measured SCS for H ions. In other words, Ni-Si behaves similar to a FEG material in the interaction with H ions. Note in this context, that these materials are known to be metallic with very low electrical resistivity (24 $\mu\Omega \cdot cm$) [61]. The electronic density-of-state (DOS) of $Ni_2Si$ shows that around the Fermi level the DOS is dominated by the d-like states of Ni [62]. The Fermi level is situated within dense occupied states, hence the material is literally a metal.

**Summary and conclusions**

We have performed a thorough examination of electronic stopping of H and He ions in Ni, Si and Ni-Si system in the regime of low velocities up to the stopping maximum. From a fundamental perspective, these energies bridges the gap between a regime, in which the electronic system of the material can be considered to be fast compared to the moving ion charge and the regime of adiabatic interaction. Due to the similarity in ion and electron velocity data as deduced in the present study represents an interesting benchmark data set for dynamical theories. The investigated energy regime is also of high technological relevance for high-resolution depth profiling techniques and ion implantation. We measured the electronic stopping cross section of these materials on supported nanometer films in backscattering geometry. Our results for protons can consistently be described by considering Ni to be a FEG material of effective electron density, since the SCS is proportional to the velocity. Furthermore, theory can perfectly predict the SCS for protons when using the electron density $r_S$ as deduced from plasmon loss experiments. For Si, DFT calculations overestimate the measured SCS by ~ 20 %, which is possibly related to the fact that Si is not a FEG material, but a semiconductor with a band gap of 1.1 eV. The deduced number of electrons per Si atoms from the plasmon experiment is about 4 e/at, whereas it is about 2 e/at in the ion beam experiments, pointing to different types of interaction. Electron-hole pair excitation appears to be more efficient in the former case due to the stronger electron-electron interactions. For He ions, deviations from the FEG model are observed in both, Ni and Si. Note, that comparable phenomena have been observed for He ions in Pt and Al. Non-adiabatic energy dissipation processes different from direct electron-hole pair excitation in binary collisions are proposed to be linked to the observations of a non-linear scaling for He as the more complex projectile. In the silicide and for He ions, we observed a clear deviation from Bragg's additivity rule: the SCS of the compound is found higher than predictions by up to 17%. This result is in contrast to data for H, for which the measured SCS of the Ni-Si alloy is virtually identical to values predicted using Bragg's rule. In general, stopping powers as predicted by DFT agrees very well with our results for H ions in Ni and in the Ni-Si silicide. Recent TD-DFT calculations exhibit the same velocity scaling of the SCS of Ni for both H and He ions as our experiments and clearly confirm that proton stopping appears to be a simpler process compared to stopping of slow He ions, for which more complex, dynamic interactions seem to contribute to the observed energy dissipation in solids.

## Acknowledgements:

Support by VR-RFI (contracts #821-2012-5144 & #2017-00646_9) and the Swedish Foundation for Strategic Research (SSF, contract RIF14-0053) supporting accelerator operation is gratefully acknowledged. Funding from the Austrian Science Fund FWF (Project No. P25704-N20) is greatly appreciated.